\documentclass{JHEP3}
\usepackage{epsfig}
\relax

\def\bes{\begin{eqnarray}}
 \def\ees{\end{eqnarray}}
\def\be{\begin{equation}}
\def\ee{\end{equation}}
\def\bs{\begin{subequations}}
\def\es{\end{subequations}}
\newcommand{\een}{\end{subequations}}
\newcommand{\ben}{\begin{subequations}}
\newcommand{\beq}{\begin{eqalignno}}
\newcommand{\eeq}{\end{eqalignno}}

\def\rt{{\tilde{r}}}
\def\ti{{\tilde{t}}}
\def\phit{{\tilde{\phi}}}
\def\ct{{\tilde{c}}}
\def\Lt{{\tilde{L}}}
\def\Dt{{\tilde{\Delta}}}
\def\chit{{\tilde{\chi}}}

\title{Holographic horizons}

\author{N. Tetradis
\\
Department of Physics, University of Athens, Zographou 157 84, Greece 

\\
{\tt ntetrad@phys.uoa.gr}}

\preprint{}

\abstract{
We examine how the (2+1)-dimensional AdS space is covered by the Fefferman-Graham system of 
coordinates for Minkowski, Rindler and static de Sitter boundary metrics. We find that, in the last two cases,
the coordinates do not cover the full AdS space. On a constant-time slice, the line delimiting 
the excluded region has endpoints at the locations of the horizons of the boundary metric. Its length, after an 
appropriate regularization, reproduces the entropy of the dual CFT on the boundary background.
The horizon can be interpreted as the holographic image of the line segment delimiting the excluded region 
in the vicinity of the boundary.}


\begin{document}

\section{AdS space in global, Poincare and Fefferman-Graham coordinates}
\label{globalcoordinates}

The AdS/CFT correspondence at low energies provides a connection between a bulk space that 
is asymptotically anti-de Sitter (AdS) and a conformal field theory (CFT)
that can be considered as living on its boundary \cite{adscft1,adscft2}.
The Fefferman-Graham system of coordinates \cite{fg} provides a very convenient parametrization of 
such as a space, starting from its boundary. In particular, it facilitates the calculation of the 
stress-energy tensor of the dual CFT \cite{Skenderis}, from which its thermodynamic properties can be deduced.
In this work we examine in detail how this coordinate system covers the bulk AdS space in the case that 
the boundary metric has a nontrivial form. In particular, we focus on Rindler and static de Sitter boundaries, which
are characterized by the presence of horizons. Our aim is to provide a pictorial analysis of this issue. 
As we switch often between various coordinate systems,
we review the ones that we employ.

The parametrization of the (2+1)-dimensional AdS space in
terms of global coordinates covers the entire manifold. 
We set the AdS length $l =1$ for simplicity throughout the paper. The metric is 
\begin{equation}
ds^2=-\cosh^2(\rt)\, d\ti^2 +d\rt^2+\sinh^2(\rt)\, d\phit^2.
\label{global} \end{equation}
The coordinate $r$ takes values in the range $0 \leq r < \infty$, while $\phi$ is periodic with values in the 
interval $-\pi < \phi \leq \pi$.
For $0\leq \ti < 2 \pi$, eq. (\ref{global}) describes AdS space. If $\ti$ is allowed to take values over the whole
real axis, we obtain the covering space of AdS. The boundary of AdS is approached for $\rt \to \infty$.
The causal structure is more visible if we define a coordinate $\chit$ through $\tan(\chit)=\sinh(\rt)$. The metric
(\ref{global}) becomes
\begin{equation}
ds^2=\frac{1}{\cos^2(\chit)} \left[ -d\ti^2 +d\chit^2+\sin^2(\chit)\, d\phit^2 \right].
\label{global2} \end{equation}
The new coordinate takes values $0\leq \chit < \pi/2$. 
The boundary is now approached for $\chit \to \pi/2$.

There is another system of coordinates, for which the AdS metric takes the form
\begin{equation}
\label{eqmetric} 
ds^2 = -r^2\,  dt^2 + \frac{dr^2}{r^2} + r^2 d\phi^2.
\end{equation}
We refer to these coordinates, which do not cover the entire AdS space,  as the Poincare parametrization 
of AdS.  The relation between the global and Poincare coordinates is \cite{review}
\begin{eqnarray}
\ti(t,r,\phi)&=&\arctan \left[ \frac{2r^2 t}{1+r^2(1+\phi^2-t^2)}\right]
\label{globalt} \\
\chit(t,r,\phi)&=&{\rm arctan} \sqrt{r^2\phi^2+\frac{\left[ 1-r^2(1-\phi^2+t^2) \right]^2}{4r^2}}
\label{globalr} \\
\phit(t,r,\phi)&=&\arctan \left[ \frac{1-r^2(1-\phi^2+t^2)}{2r^2 \phi} \right].
\label{globalf} \end{eqnarray}
A thorough analysis of the part of AdS covered by the Poincare coordinates is given in \cite{poincare}.
As the spatial part of the AdS boundary is compact, the global coordinate $\phit$ is
periodic with period $2\pi$. It is obvious then from eq. (\ref{globalf}) that the limits $\phi\to\pm\infty$ of the 
Poincare coordinate $\phi$ must be identified. The slice $t=\ti=0$ is covered entirely 
by both coordinate systems.

The trivial redefinition $z=1/r$ puts the metric (\ref{eqmetric}) in the form
\begin{equation}
\label{eqmetric2} 
ds^2 = \frac{1}{z^2}\left(  dz^2- dt^2 + d\phi^2 \right).
\end{equation}
This is the simplest example of the Fefferman-Graham parametrization \cite{fg} of asymptotically AdS spaces, 
with the conformal boundary 
located at $z=0$. The general form of the metric for such spaces can be obtained as an expansion around $z=0$.
(However, the global properties of the geometry can be quite nontrivial \cite{global}.) 
We are interested in (2+1)-dimensional metrics that satisfy Einstein's equations with 
a negative cosmological constant. All these are locally isomorphic to AdS space and have the form \cite{finite}
\be\label{eq2} ds^2 = \frac{1}{z^2} \left[ dz^2 + g_{\mu\nu} dx^\mu dx^\nu \right], \ee
where
\be g_{\mu\nu} = g_{\mu\nu}^{(0)} + z^2 g_{\mu\nu}^{(2)} 
+ z^4 g_{\mu\nu}^{(4)}. \label{fefg} \ee
The metric of the conformal boundary is $g_{\mu\nu}^{(0)}$. In the example of eq. (\ref{eqmetric2}) we have
a flat boundary with $g_{\mu\nu}^{(0)}=\eta_{\mu\nu}$ and an identification of the limits $\phi\to\pm \infty$ for every value of
$t$. 

In the following we discuss particular cases of eq. (\ref{eq2}), (\ref{fefg}) with Rindler or de Sitter boundary metrics. 
We are interested in static properties of such space-times. For this reason we focus on a constant-time slice, which we choose
as the one with $t=0$. The corresponding value in global coordinates is $\ti=0$. Both the Poincare and global systems of coordinates
cover entirely this slice, while the Fefferman-Graham system covers only parts of it. Our main observation is that the missing parts are
a consequence of the presence of horizons in the boundary metric. 

\begin{figure}[t]
\begin{center}
\epsfig{file=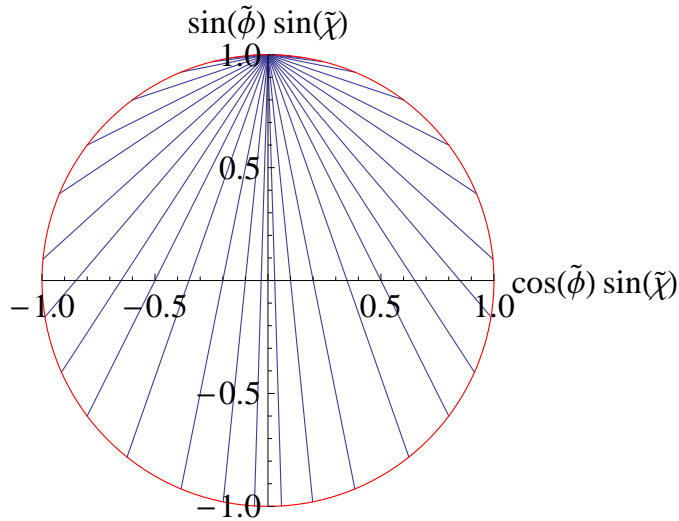,height=7cm}
\end{center}
\caption{Lines of constant $\phi$ for a Minkowski boundary.}
\label{minkowski}
\end{figure}

As a starting point we depict in fig. \ref{minkowski} how the slice $t=\ti=0$ is covered by lines characterized by a constant 
value of the spatial coordinate on the conformal boundary and the ``bulk" coordinate $z$ varying over its whole allowed range:
$0\leq z < \infty$. In the case (\ref{eqmetric2}) with a flat boundary, each line corresonds to a constant value of $\phi$ and 
varying $z$. The corresponding value of the global coordinates $\phit$, $\chit$ can be obtained through the transformation
(\ref{globalr}), (\ref{globalf}). The AdS boundary is located at $\chit=\pi/2$, which corresponds to the bounding circle in
fig. \ref{minkowski}. Each line starts with $z=0$ on the boundary and terminates again on the boundary when $z\to \infty$.
All lines terminate at the point $\phit=\chit=\pi/2$. The values of $\phi$ for each line increase as the starting point 
moves counterclockwise around the bounding circle. The lines with $\phi\to-\infty$ start on the left of the point 
$\phit=\chit=\pi/2$ and infinitesimally close to it.  The lines with $\phi\to \infty$ start on the right of this point 
and infinitesimally close to it. This feature is consistent with the identification of the limits $\phi\to\pm\infty$ that we
mentioned earlier.

\section{Rindler boundary}
\label{rindlerboundary}

We would like to put the (2+1)-dimensional AdS metric in a form such that the conformal boundary is of the Rindler type.
This can be achieved for the Rindler wedge $(x>0)$. The metric (\ref{eqmetric}) can be written as
\begin{equation}
ds^2
= \frac{1}{z^2} \left[ dz^2 
- a^2 x^2 \left(1+\frac{z^2}{4x^2} \right)^2 dt^2 
+  \left(1-\frac{z^2}{4x^2} \right)^2 dx^2 \right].
\label{rindb}
\end{equation}
This is a particular case of the Fefferman-Graham parametrization (\ref{fefg}),
characterized by a Rindler boundary
at $z=0$ with metric $g_{\mu\nu}^{(0)}dx^\mu dx^\nu=-a^2 x^2 dt^2+dx^2$ .
The coordinate transformation that achieves this does not affect the time coordinate. It is given by
\begin{eqnarray}
r(z,x)&=&a\left(\frac{x}{z}+\frac{z}{4x} \right)
\label{bbb22} \\
\phi(z,x)&=&\frac{1}{a}\log\left[a x \right]-\frac{8}{a(4+z^2/x^2)}.
\label{bbb33}
\end{eqnarray}
Notice that the transformation maps the whole region near negative infinity 
for $\phi$ to the neighborhood of zero for $x$, with $x>0$.

\begin{figure}[t]
\begin{center}
\epsfig{file=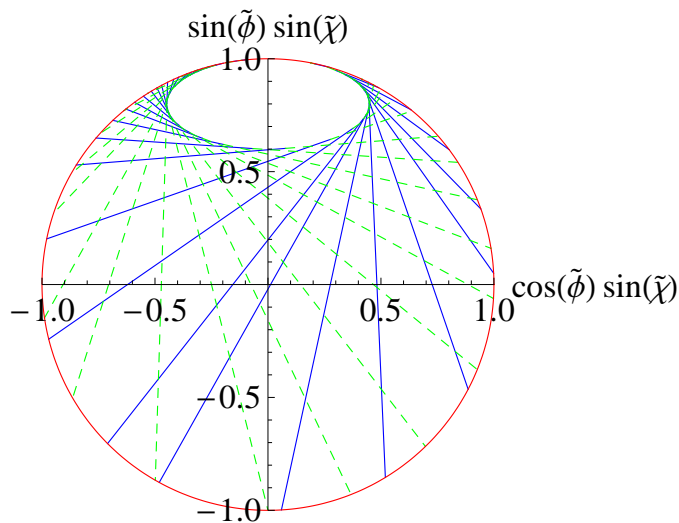,height=7cm}
\end{center}
\caption{Lines of constant $x$ for a Rindler boundary with $a=0.5$.}
\label{rindler1}
\end{figure}

A crucial property of the transformation (\ref{bbb22}), (\ref{bbb33}) is that,
for fixed $x$,  the function $r(z,x)$ has a minimal value at the point where 
$\partial r(z,x)/\partial z=0$. This relation generates the line
\be
z_t(x)=2x,
\label{zmx} 
\ee
on which 
\be
r_t(x)=a.
\label{rmx} 
\ee
The corresponding values of $\phi$ are 
\be
\phi_t(x)\equiv\phi(z_t(x),x)=(\log[ax]-1)/a.
\label{phimx} 
\ee
We have seen that $r(z,x)$ has a constant minimal value. 
As a result the slice $t=0$ is not fully covered by the Fefferman-Graham system of coordinates. 
We demonstrate this fact in fig. \ref{rindler1}, in which we depict the form of the lines with constant 
$x$ and $0\leq z\leq z_t(x)$ for $a=0.5$. They are the solid line segments that start at the boundary and finish at some point in the interior
of AdS space. The values of $x$ for each line increase as the starting point 
moves counterclockwise around the bounding circle. The lines with $x\to 0$ start on the left of the point 
$\phit=\chit=\pi/2$ and infinitesimally close to it.  The lines with $x\to \infty$ start on the right of this point 
and infinitesimally close to it. From eq. (\ref{bbb33}) we obtain $\phi(0,x)=(\log[ax]-2)/a$. On the boundary,
the limits $x\to 0$ and
$x\to \infty$ correspond to $\phi\to-\infty$ and $\phi\to\infty$, respectively. The identification of the latter two results in
the identification of the first two limits as well. 

The extensions of the lines when $z$ is allowed to take values $z_t(x) \leq z <\infty$ for given $x$ 
cover the AdS space for a second time, leaving the 
same empty region. We demostrate this feature in fig. \ref{rindler1}. Each solid line segment (corresponding to 
 $0\leq z\leq z_t(x))$ has a dashed extension (corresponding to $z_t(x) \leq z <\infty$) that ends on the boundary.
In this way, the system of coordinates covers the AdS space twice, leaving an empty region whose boundary 
corresponds to $r_t(x)=a$. The line on which the Fefferman-Graham system of coordinates turns around 
and covers the AdS space for a second time is determined by
$\partial r(z,x)/\partial x=0$. This relation gives again eq. (\ref{zmx}), which resulted from
$\partial r(z,x)/\partial z=0$. This coincidence is particular to the Rindler case. In the following
section we shall consider the case of a de Sitter boundary, for which the various partial derivatives do not vanish
along the same line.

\begin{figure}[t]
\begin{center}
\epsfig{file=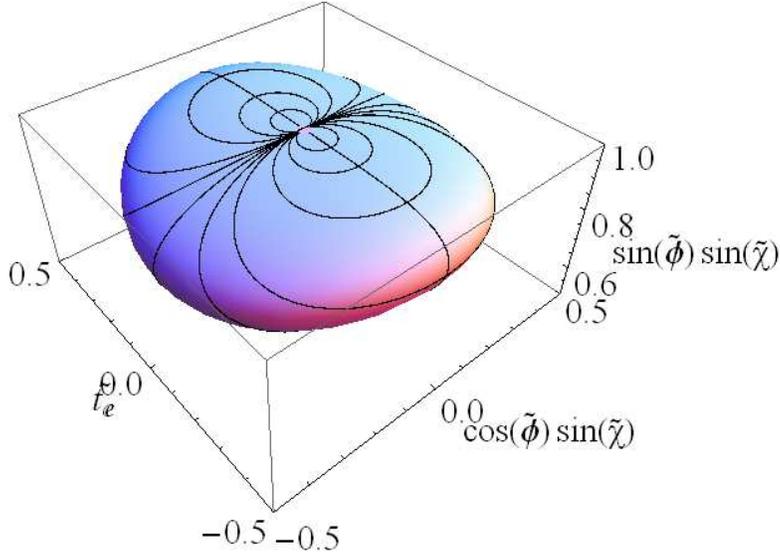,height=8.2cm}
\end{center}
\caption{The region not covered by Fefferman-Graham coordinates for a Rindler boundary with $a=0.5$.}
\label{rindler2}
\end{figure}

In order to examine the form of the excluded region for $t\not= 0$, it is more convenient to perform a 
Euclidean rotation of the time coordinates appearing in the metrics (\ref{global2}) and
(\ref{eqmetric}):  $\ti_E=-i\,\ti$, $t_E=-i\,t$. The global and Poincare coordinates are related through the 
transformation 
\begin{eqnarray}
\ti_E(t_E,r,\phi)&=&{\rm arctanh} \left[ \frac{2r^2 t_E}{1+r^2(1+\phi^2+t_E^2)}\right]
\label{globalte} \\
\chit(t_E,r,\phi)&=&{\rm arctan} \sqrt{r^2\phi^2+\frac{\left[ 1-r^2(1-\phi^2-t_E^2) \right]^2}{4r^2}}
\label{globaler} \\
\phit(t_E,r,\phi)&=&\arctan \left[ \frac{1-r^2(1-\phi^2-t_E^2)}{2r^2 \phi} \right].
\label{globalfe} \end{eqnarray}
The Euclidean AdS is covered completely by both systems \cite{review}. 
The Fefferman-Graham coordinates that we consider in this work result from the Poincare ones through transformations that do not
involve the time coordinate.  For static boundaries, the use of Euclidean time is expected to give 
a description of field theories on AdS equivalent to that based on Lorentzian time. 

In fig. \ref{rindler2} we present the part of the Euclidean AdS that is not covered by Fefferman-Graham coordinates with
a Rindler conformal boundary. The boundary of AdS (not shown in fig. \ref{rindler2}) 
is the cylinder of radius equal to 1 that results from the continuation of the bounding
circle in fig. \ref{rindler1} along the Euclidean time direction. In fig. \ref{rindler2} 
we depict the boundary of the excluded volume, which corresponds to the surface on which the Poincare coordinate $r$
obtains its minimal value
$r=a=0.5$ . The closed circular lines on this surface have constant Poincare time $t_E$, while
$\phi$ varies from $-\infty$ to $\infty$ along them. 
All these lines start and terminate at the point ($\ti_E=0,\phit=\pi/2,\chit=\pi/2$) on the boundary of AdS. 
Each line determines the narrowest part of a throat connecting the
two asymptotic regions, at $z\to 0$ and $z\to \infty$, that we discussed 
in the context of fig. \ref{rindler1} for $t_E=0$. In a following section
we shall examine the conjecture that the length of each line is related to the 
entropy of the dual CFT on a (1+1)-dimensional Rindler background. 

According to the AdS/CFT correspondence, the (2+1)-dimensional metric of the form 
(\ref{eq2}), (\ref{fefg})  encodes information on the dual CFT on a gravitational background 
given by $ g_{\mu\nu}^{(0)} $. In 2+1 dimensions the Fefferman-Graham expansion terminates at 
the term $\sim z^4$. As a result we can obtain a second interpretation in terms of the dual CFT on a background with 
metric $g_{\mu\nu}^{(4)} $. This possibility is apparent through the simple coordinate change $z=1/z'$. 
The limit $z'\to 0$ is equivalent to the limit $z\to \infty$ that leads back to the AdS boundary, as we have seen
in the example of the Rindler boundary. 
The Rindler case is particularly simple because the background $ g_{\mu\nu}^{(4)} $ is again of the Rindler type, 
as can be verified through the coordinate change $x=1/(4x')$. 
We can view the (part of the) AdS space covered by the Fefferman-Graham coordinates with $0\leq x <\infty$ and
$0\leq z \leq z_t(x)$ as providing a holographic description of the dual CFT on the Rindler background at $z=0$. 
The second copy of (part of) the AdS space covered by  $0\leq x <\infty$ and
$z_t(x)\leq z \leq \infty$ (where the last equality corresponds to the point on the AdS boundary obtained for
$z\to \infty$) describes the dual CFT on a Rindler background at $z'=0$. 
In the holographic description of both dual theories, part of the AdS space is not covered by the Fefferman-Graham system of coordinates. 

\section{Static de Sitter boundary}
\label{sdsb}

We now turn to the case of a de Sitter boundary in static coordinates.
The metric (\ref{eqmetric}) can be put in the form
\begin{eqnarray}
ds^2
= \frac{1}{z^2} \Bigg[ dz^2 
&-& (1-H^2\rho^2)\left(1+\frac{1}{4}\left[\frac{H^2}{1-H^2\rho^2}-H^2\right]z^2 \right)^2 dt^2 
\Biggr.
\nonumber \\
\Biggl.
&+& \left(1-\frac{1}{4}\left[\frac{H^2}{1-H^2\rho^2}+H^2\right]z^2 \right)^2  \frac{d\rho^2}{1-H^2\rho^2} \Biggr],
\label{eqmetric4} \end{eqnarray}
with a de Sitter boundary at $z=0$ with metric $g_{\mu\nu}^{(0)}dx^\mu dx^\nu=-(1-H^2\rho^2)dt^2+d\rho^2/(1-H^2\rho^2)$.
We consider the causally connected region $-1/H < \rho < 1/H$. 
The transformation between Poincare and Fefferman-Graham coordinates does not affect the time coordinate. It is given by 
\begin{eqnarray}
r(z,\rho)&=&\frac{\sqrt{1-H^2\rho^2}}{z}+\frac{H^4\rho^2}{4\sqrt{1-H^2\rho^2}}z
\label{bbb2} \\
\phi(z,\rho)&=&\frac{1}{2H}\log\left[\frac{1+H\rho}{1-H\rho} \right]
-\frac{H^2\rho z^2}{2(1-H^2\rho^2+H^4\rho^2z^2/4)}.
\label{bbb3}
\end{eqnarray}
 It is noteworthy that the transformation 
maps the region near $-\infty$  for $\phi$ to the vicinity of $-1/H$ for $\rho$, and the region near $\infty$ for
$\phi$ to the vicinity of $1/H$ for $\rho$. The identification of the limits $\phi\to\pm\infty$  results in the 
identification of the limits $\rho\to\pm 1/H$ as well.

\begin{figure}[t]
\begin{center}
\epsfig{file=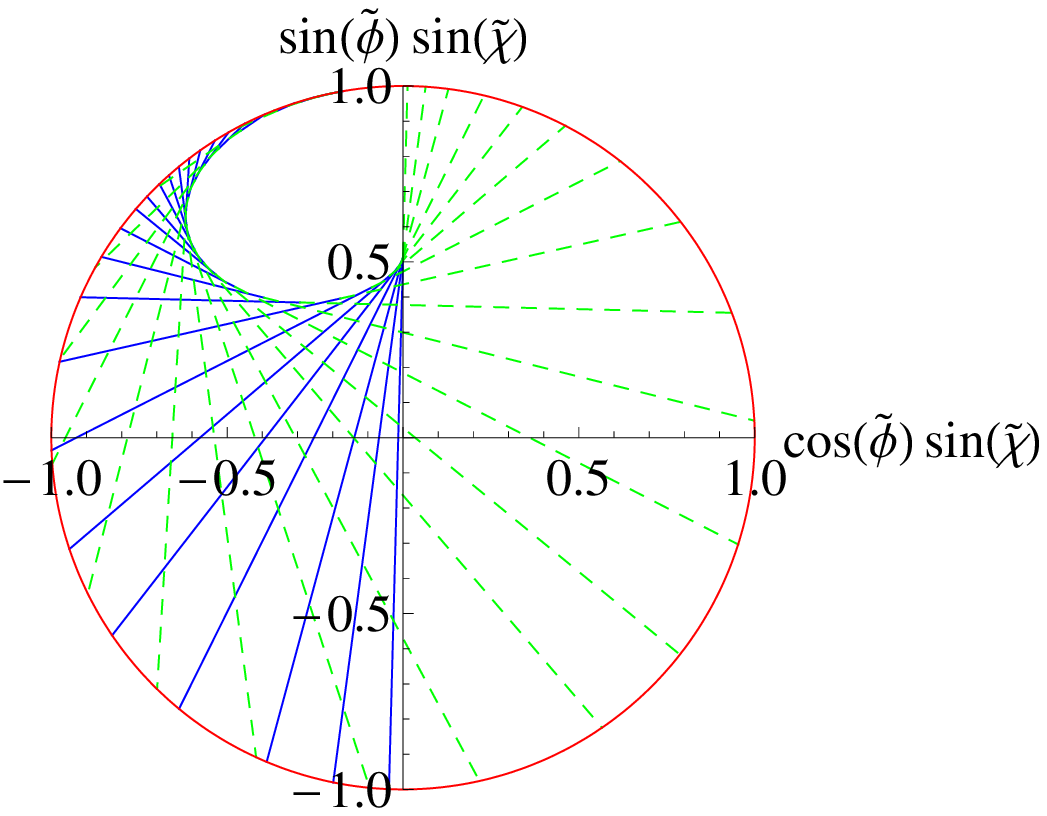,height=7cm}
\end{center}
\caption{Lines of constant $\rho<0$ for a static de Sitter boundary with $H=0.8$.}
\label{desitter1}
\end{figure}

\begin{figure}[t]
\begin{center}
\epsfig{file=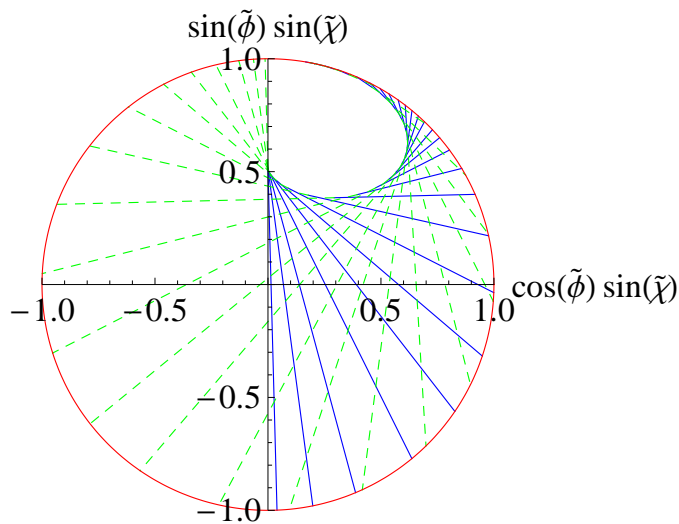,height=7cm}
\end{center}
\caption{Lines of constant $\rho>0$ for a static de Sitter boundary with $H=0.8$.}
\label{desitter2}
\end{figure}

The way in which the AdS space is covered by the Fefferman-Graham coordinates for a de Sitter boundary is more complicated that in
the Rindler case. We concentrate on the $t=\ti=0$ slice.
In figs. \ref{desitter1}, \ref{desitter2} we depict the form of the lines with constant positive or negative 
$\rho$, respectively, and $0\leq z<\infty$ for
$H=0.8$. In fig. \ref{desitter1}
the values of $\rho$ for each line increase as the starting point 
moves counterclockwise around the bounding circle. The solid part of each line starts from the AdS boundary, with the initial
point corresponding to $z=0$. The dashed part is obtained for large values of $z$, with the boundary being approached 
again for $z\to\infty$. 
The lines with $\rho \to -1/H$ start on the left of the point 
$\phit=\chit=\pi/2$ and infinitesimally close to it. The starting points (corresponding to $z=0$) cover half of the AdS boundary for
$\rho$ taking values $-1/H\leq \rho \leq 0$. On the other hand, the endpoints (obtained for $z\to \infty$) cover the whole AdS boundary.

For the transition from the solid to the dashed part of each line we selected the point at which the system of coordinates
starts covering (part of) the AdS space for a second time. This point is given by the solution of
$\partial r(z,\rho)/\partial \rho=0$, which is
\be
z_t(\rho)=\frac{2}{H}\sqrt{\frac{1-H^2\rho^2}{2-H^2\rho^2}}.
\label{zmrho1} \ee
It corresponds to 
\begin{eqnarray}
r_t(\rho)\equiv r(z_t(\rho),\rho)&=&\frac{H}{\sqrt{2-H^2\rho^2}},
\label{rmrho}
\\
\phi_t(\rho)&=&\frac{1}{2H}\log\left[\frac{1+H\rho}{1-H\rho} \right]
-\rho.
\label{phimrho}
\end{eqnarray}
Notice that the minimal value of $r(z,\rho)$ as a function of $z$ for given $\rho$ is obtained for $\partial r(z,\rho)/\partial z=0$. 
This gives 
\be
z_m(\rho)=2\sqrt{\frac{1-H^2\rho^2}{H^4 \rho^2 }},
\label{zmrho}
\ee
which satisfies $z_m(\rho)\geq z_t(\rho)$ for all $\rho$ and $z_m(\rho)/z_t(\rho)\to 1$ for $\rho\to\pm 1$.

The most interesting feature of fig. \ref{desitter1} is that part of the AdS slice is not covered by the coordinates.
The boundary of the excluded part is not determined in general by a constant value of the Poincare coordinate $r$, as in the Rindler case.
Only near the point $\phit=\chit=\pi/2$ (top of the diagram), where  $\rho\simeq -1/H$, 
the boundary of the excluded part coincides with the line of constant Poincare coordinate $r=r_t(|\rho|\simeq1)\simeq H$. 
This means that this region maps the vicinity of one of the horizons of the (1+1)-dimensional de Sitter geometry.  

In fig. \ref{desitter2} we depict the lines corresponding to $0\leq \rho \leq 1/H$. The solid parts of these lines start from points
on the AdS boundary obtained for $z=0$. The starting points cover the second half of the AdS boundary. On the other hand,
the endpoints of the lines at the end of the dashed segments, obtained for $z\to \infty$, cover the full AdS boundary. 
There is a part of the AdS slice that is not covered by the
coordinates. Near the top of the diagram, the boundary of the excluded part coincides with the line of constant 
$r=r_m(\rho\simeq 1)\simeq H$.  
This region maps the vicinity of the second horizon of the (1+1)-dimensional geometry.  

Similarly to the Rindler case, 
we view the (part of the) AdS space covered by the Fefferman-Graham coordinates with $-1 \leq \rho \leq 1$ and
$0\leq z \leq z_t(\rho)$ as providing a holographic description of the dual CFT on the static de Sitter background at $z=0$. 
The second copy of (part of) the AdS space covered by  $-1 \leq \rho \leq 1$ and
$z_t(\rho)\leq \rho \leq \infty$ (where the last equality corresponds to the point on the AdS boundary obtained for
$z\to \infty$) describes the dual CFT on a different background at $z'=1/z=0$. Contrary to the Rindler case, this second 
background does not have an obvious physical interpretation. For this reason we concentrate on the dual theory 
at $z=0$. On the gravity side, we consider only the part of AdS covered by the solid lines in figs. \ref{desitter1}, \ref{desitter2}.

The extension of this picture to the full AdS geometry is too complicated to be represented pictorially. 
However, for static boundary metrics, such as the ones we are considering, the determination of thermodynamic properties
of the dual CFT can be performed at any time. We shall confirm this fact for the Rindler entropy in the next section.
On the other hand, for the de Sitter entropy we shall consider only the slice $t=\ti=0$.

\section{Horizons and entropy}
\label{lthr}

The proposal of ref. \cite{tetradis} is that 
the entropy associated with the dual theory on a nontrivial background is related to the part of the
AdS space not covered by the Fefferman-Graham system of coordinates. More specifically, the entropy at a given
time is assumed to be proportional to the length $A$ of the line defining the boundary of the part of AdS not
covered by the coordinates.
The exact relation between the entropy $S$ and the length $A$ is 
\be
S= \frac{1}{4G_3} A,
\label{entropyy} \ee
with $G_3$ Newton's constant of the (2+1)-dimensional theory. In the following we review how this expression
reproduces the Rindler and de Sitter entropies. The detailed calculation has been presented in ref. \cite{tetradis}, but
several related issues are clarified through inspection of 
figs. \ref{rindler1}-\ref{desitter2}.

For a Rindler boundary at $z=0$, the part of AdS space not covered by the coordinates (\ref{rindb}) is delimited by
the line of constant $r$ given by eq. (\ref{rmx}) .  For  $t=0$ the omitted region is depicted in fig. \ref{rindler1}.
The corresponding values of $z$ and $\phi$ are given by 
eqs. (\ref{zmx}, (\ref{phimx}), respectively. The entropy can be expressed as a line integral, with various forms depending on
the integration variable:
\be
S=\frac{1}{4G_3}  \int_{-\infty}^{\infty} a d\phi = \frac{1}{4 G_3} \int_0^\infty \frac{dx}{x} = 
 \frac{1}{4 G_3} \int_0^\infty \frac{dz}{z} .
\label{srind} \ee
All the above integrals are infinite.
The divergences arise from the two ends of the line that approach the point $\phit=\chit=\pi/2$ (top of the diagram of fig. \ref{rindler1}). 
The last integral in eq. (\ref{srind}) demonstrates that the divergence is related to the infinite volume near the 
boundary of AdS space. 
The way to handle this situation is to introduce a cutoff $\epsilon$ for 
the line element, with $\epsilon \ll 1$. The dominant contribution to the integral of eq. (\ref{srind}) can be written
as 
\be
S=\frac{2}{4 G_3} \int_{\epsilon} \frac{dz}{z},
\label{regulr} \ee
with the factor of 2 arising from the two limits $\phi\to\pm \infty$.
We have not introduced an explicit upper limit, as its exact value is
irrelevant for $\epsilon \to 0$. 

In order to extract the physical meaning of eq. (\ref{regulr}), we must compare it with the effective Newton's constant $G_2$
of the (1+1)-dimensional theory. For a theory living on the $z=0$ boundary, $G_2$ is given by the expression.
\be
\frac{1}{G_2}=\frac{1}{G_3}\int_{\epsilon}\frac{dz}{z}.
\label{rsgn2} \ee
The cutoff $\epsilon$ is again introduced in order to regulate a divergence generated by the infinite AdS volume near the boundary.
We have omitted the upper limit, as the integral is dominated by the lower one for $\epsilon \to 0$.
The fact that $G_2$ tends to zero in the same limit is consistent with the absence of dynamics for the 
boundary gravity in the context of AdS/CFT.

Comparison of eqs. (\ref{regulr}), (\ref{rsgn2}) indicates that we can interpret eq. (\ref{srind}) as
\be
S=\frac{2}{4G_2}.
\label{srindd} \ee
Repeating the calculation of ref. \cite{laflamme} for the two-dimensional case results in an expression 
for the entropy of the two-dimensional Rindler wedge that is half the value given by eq. (\ref{srindd}). 
The discrepancy is related to the symmetry around the vertical axis of  the region not covered by the 
Fefferman-Graham coordinates in fig. \ref{rindler1}.
The origin of this feature 
can be traced to the periodicity of the global coordinate $\phit$ arising 
from the compactness of the spatial direction on the AdS boundary.
As a result, in the construction with a Rindler conformal boundary, the limits $x\to 0$ and $x \to \infty$ of the 
Fefferman-Graham coordinate $x$ must be identified. 
The two limits correspond to the two 
regions on either side of the point $\phit=\chit=\pi/2$ in fig. \ref{rindler1}, which are completely symmetric. 
The region on the left provides a
holographic description of the vicinity of the Rindler horizon at $x=0$. The region on the right has an identical form, 
mimicking
the presence of a horizon at $x\to \infty$.
We are thus led to the conclusion that the construction of section \ref{rindlerboundary} does not represent the conventional Rindler
space. It provides a holographic description of the Rindler wedge ($0\leq x < \infty$), 
with an identification of the limits $x\to 0$ and $x\to \infty$. 

The above considerations indicate that the entropy of the conventional Rindler space is associated only with 
the region near $x=0$, which provides an acceptable holographic description of the 
vicinity of the Rindler horizon. The defining feature is the absence of complete coverage of the $t=0$ slice of AdS space. The
entropy can be identified with the length of the boundary of the excluded region. This is dominated by the
part of the line near the AdS boundary, and is given by 
\be
S_R= \frac{1}{4G_2}.
\label{srinddd} \ee
The extension of this feature to different time slices is given in fig. \ref{rindler2}. All slices of constant 
Euclidean time $t_E$ (closed circular lines in fig. \ref{rindler2}) go through the point $\phit=\chit=\pi/2$ on the
AdS boundary. Repeating the above analysis for anyone of them will lead to eq. (\ref{srinddd}), as the 
expression for the entropy is dominated by the divergent integral near the AdS boundary.

The case of a de Sitter boundary has a more straightforward interpretation. 
The identification of the limits $\rho \to \pm 1/H$, imposed by the periodicity of the global
coordinate $\phit$, does not alter the physical picture, because the two limits correspond to the 
two symmetric horizons of the (1+1)-dimensional de Sitter space. The part of the AdS space 
covered by the solid lines in figs. \ref{desitter1}, \ref{desitter2} provides a holographic description of
the causally connected region between the horizons. 
The horizons and their vicinity are mapped on the boundary of the region covered  by the Fefferman-Graham coordinates in the
neighborhood of the point $\phit=\chit=\pi/2$.
The sum of the lengths of the two lines delimiting this region and ending at the point  $\phit=\chit=\pi/2$
gives the de Sitter entropy. 
The entropy is given by eq. (\ref{regulr}), which results in  
\be
S_{dS}=\frac{1}{2G_2}.
\label{srdes} \ee
This is the correct expression for the entropy of two-dimensional de Sitter space \cite{dsen}.

Based on the above, we can conclude that, 
in the context of the AdS/CFT correspondence, it seems natural to attribute the entropy of the dual theory 
on a certain boundary metric to
the presence of a part of the AdS space that is not covered by the Fefferman-Graham coordinates.
These have a special status, at they are adopted to a boundary observer and provide a natural
holographic interpretion of the bulk space. The clearest example of the
connection between the boundary entropy and the lack of coverage of the AdS space by the coordinates
involves the BTZ black hole in the bulk with a Minkowski conformal boundary \cite{tetradis}. 
The examples of a Rindler and de Sitter boundary that
we discussed in detail provide additional support for this connection. The implications for a configuration with an
AdS black hole in the bulk and an expanding boundary have been discussed in refs. 
\cite{ast,tetradisold,lamprou,tetradis}.

From the holographic perspective, the bulk space may be viewed as a represenation of the boundary CFT, 
with the extra dimension playing the role of the energy scale. 
In the examples we discussed the bulk is locally isomorphic to AdS space, while for certain
boundary metrics there are bulk regions that are not covered by the coordinates.  
A bulk field with an arbitrary dependence on
the Fefferman-Graham coordinates will not take values in these AdS regions.
Our suggestion is that, since
this part of the bulk is not included in the construction of the dual theory, it must be associated with the
entropy of the CFT.
This definition of entropy bears strong similarity to the entropy generated by quantum entanglement between classically 
disconnected regions of space. 

For Rindler and de Sitter bounaries, a horizon of the boundary metric plays a special role. It defines a point
separating two causally disconnected regions of the boundary. It also determines the beginning of the line delimiting
the part of the bulk excluded by the Fefferman-Graham parametrization. We can view the horizon as the holographic
image of the part of this line in the near-boundary region. The entropy is proportional either to area of the horizon, or the 
area of its bulk extension. The two areas  have different dimensionalities. However, the proportionallity constants in the 
two cases (involving the effective or the bulk Newtons's constant, respectively) compensate for this difference.

As a final comment, we point out that the holographic interpretation of entropy is not necessarily bound to the presence of a horizon.  
For a BTZ bulk and a Minkowski boundary, there is no horizon on the boundary.
However, the connection with the entropy 
persists: The excluded part of the bulk lies away from the boundary and is
related to the thermal entropy of the dual CFT \cite{tetradis}.

\section*{Acknowledgments}
I wish to thank T. Christodoulakis, G. Diamandis, E. Kiritsis, I. Papadimitriou, K. Skenderis, P. Terzis
for useful discussions. 
This work was supported in part by the EU Marie Curie Network ``UniverseNet'' 
(MRTN--CT--2006--035863) and the ITN network
``UNILHC'' (PITN-GA-2009-237920).

\end{document}